\documentclass[aps,prb,preprint,groupedaddress]{revtex4-1}

\usepackage{graphicx} 
\usepackage{subfigure}
\usepackage{amsmath,amsfonts,amssymb,latexsym}
\usepackage{color}
\usepackage{natbib}
\usepackage[utf8]{inputenc}
\usepackage[T1]{fontenc}
\usepackage[english]{babel}
\usepackage{float}
\usepackage{braket}
\usepackage{mciteplus}
\def\url#1{}

\begin{document}

\title
{Correlation functions in electron-electron and
electron-hole double quantum wells: temperature, density and barrier-width dependence.}

\author
{M.W.C. Dharma-wardana}\email[Email address:\ ]
{chandre.dharma-wardana@nrc-cnrc.gc.ca}
\affiliation{
National Research Council of Canada, Ottawa, Canada, K1A 0R6
}

\author{D. Neilson and F. M. Peeters}
\affiliation{ 
Department of Physics, University of Antwerp, Groenenborgerlaan 171,  2020 Antwerp, Belgium
}

\begin{abstract}
The classical-map hyper-netted-chain (CHNC) scheme, developed for treating
fermion fluids at strong coupling and at finite temperatures, 
is applied to electron-electron and electron-hole double quantum wells.  
The pair distribution functions  and the local field factors needed in linear
 response theory are determined 
for a range of temperatures, carrier densities, and barrier widths typical 
for experimental double quantum well systems in GaAs-GaAlAs.
For electron-hole double quantum wells, a large enhancement in the pair distribution
 functions is found for small carrier separations. 
The CHNC equations for electron-hole systems no longer hold at low densities where
bound-state formation occurs. 
\end{abstract}
%
%\pacs{\textcolor{red}{\bf xyz}}

\maketitle

\section{Introduction}

Quantum nanostructures with charge carriers confined in reduced
 dimensions\cite{afs} continue to be of
great interest.   Enormous progress in fabrication
techniques has realized systems in which the carriers have 
extremely high mobilities and can be taken down to very
 low densities\cite{Zhu2003,Yoon1999}.  
A system consisting of a pair of  strongly coupled quasi two-dimensional (2D) layers 
of mutually interacting electron or hole fluids separated by
a thin insulating layer with negligible tunneling, is predicted  to
support novel phases stabilized by interlayer Coulomb interactions.  These phases include 
excitonic superfluids\cite{Butov1994,Timofeev2000,Cheng1995,Kono1997,Marlow1999,Filinov2003,Perali2013}, 
coupled Wigner crystals and charge density waves\cite{Swi91,Zarenia2017}, 
and entangled states relevant in electronics and quantum information devices\cite{Jakubczyk17}.  
Coupled double layer systems can be fabricated in conventional 
semiconductor heterostructures using two adjacent quantum 
wells\cite{Sivan1992,Kane1994,Butov1994,Keogh2005,Croxall2008,Seamons2009,ZhengAPL16} or, alternatively,  
they can be fabricated using two sheets of atomically thin materials like monolayer or bilayer graphene,
separated by a high insulating barrier of hBN or WSe$_2$ \cite{Gorbachev2012,Li2016,Lee2016,Burg18}.   

Coupled double layer systems,  which can be represented as coupled 2D interacting plasmas, provide a means of
studying  intricate many-particle interactions that depend on carrier density, masses, spin, as well as temperature. 
At low densities and for small separations of  the layers, the carrier correlations can become very strong, especially for coupled electron-hole layers with their attractive interactions. 
\'{S}wierkowski {\it et al.}\cite{Swierkowski1995} demonstrated the importance of 
electron-hole correlations in experimental electron-hole drag resistivity data.      
Correlations in double quantum wells have been studied using quantum Monte Carlo 
simulations\cite{DePalo02,Senatore2003,Filinov2003,Filinov2004,Maezono2013,LopezRios2018}.

At finite temperatures, the degeneracy of the carriers is controlled  by the ratio of their temperature
to the Fermi energy, $t=T/E_F$.  When $t>1$,  degeneracy starts to significantly decrease, and this is 
accompanied by a decreasing importance of quantum effects, which in turn affects the correlations.  
For low-density holes in GaAs with their relatively large effective mass, the Fermi temperature 
can be as little as a few Kelvin.
For example, for a hole layer density $n=4\times 10^{10}$ cm$^{-2}$ in GaAs, 
the Fermi temperature is only $4$ K.   Hence the 
need to account for the  temperature dependence of  exchange and correlation among carriers 
becomes unavoidable even at nominally ``low'' temperatures, and this is an over-arching objective of this study.  
 
The direct evaluation of pair distribution functions and linear response functions 
of quantum systems is extremely important, since all static properties (e.g., thermodynamics) as
 well as linear response properties (e.g., conductivities) of a system can be accessed
if the corresponding pair distribution functions are known, {\em without} recourse to the
many-body wavefunction \cite{apvmm}.

In this study we calculate
 the temperature dependent pair distribution functions and local field factors 
for electron-electron (e-e) and electron-hole (e-h) double quantum
 wells  that are 
needed for finite temperature studies such as in the calculation of drag resistance,  
plasmon dispersions, hot electron relaxation, as well as for the calculation of thermodynamic properties   
\cite{Lozovik1975,Lozovik1976,Lozovik1976a,ZhengAPL16,Littlewood1996,Lozovik1997}.
However,  because of its intrinsic importance as well as for simplicity we restrict
 ourselves in this study to
symmetric double layer systems, where we  consider equal densities and equal effective
masses of carriers in both layers. Here we note that Maezono {\it et al.} \cite{Maezono2013}
who studied excitonic condensation at zero temperature  have
followed the same philosophy and  state that ``we have studied the
simplest possible such model system, with equal electron
and hole populations and equal masses, and parallel infinitely thin two-dimensional
 layers of variable separation and carrier density. It is important to establish
 the behavior of
this simple system before more complicated cases such as
those of unequal electron and hole masses [8] and/or unequal
electron and hole densities can be tackled with confidence''.
Such symmetric systems describe a very important class of double quantum
 wells manifested by graphene-like bilayers where a variety of effects
 arise~\cite{Perali2013, Zarenia2017, Gorbachev2012,Li2016,Lee2016,Burg18,Zarenia18,DePalo02,Maezono2013}.
 A further justification is that the  symmetric system  is likely to be the case
 for which quantum Monte Carlo
 and Feynman-path integral methods are likely to
be feasible for providing bench marks \cite{Maezono2013} for
 finite temperature systems.
 Since no external field is applied, we consider the unpolarized case in this study. 
Negligible tunneling of carriers across the insulating barrier separating the layers is assumed.

In stochastic methods like Quantum Monte Carlo (QMC) simulations\cite{DePalo02} 
the explicit many-body wave function has to be used, which limits this method to 
a small number of  carriers (typically $N\sim$ 100). If there are two types of
 carries (e.g., two wells) with two types of spin, a QMC calculation with $\sim$100
 particles implies that there are perhaps $\sim 24$ particles per species, and
 the statistical errors become important unless larger simulations are possible.
 Given the sensitivity of the calculations
to the assumed form of the wavefunction, boundary conditions, backflow effects etc., 
reliable calculations at finite-$T$ may remain a challange.
Unfortunately, alternative perturbation methods based on Feynman graphs,
 quantum kinetic
 equations etc.,  are either limited to weak-coupling approximations or
 ``decoupling approximations'' \cite{stls}.  
Such kinetic equation methods often fail to even obtain non-negative pair distribution
 functions $g({r})$, an elementary {\it a priori} requirement since $g({r})$ is the probability,
 given a particle at the origin, of finding another particle at distance ${r}$ from the origin. 
 
The classical-map hyper-netted chain (CHNC) method introduced in Ref.\ \onlinecite{prl1},  
uses a mapping of the quantum electron system
 to an ``equivalent'' classical electron system, and is able to directly evaluate pair distribution functions and
linear response functions of the quantum system.  It has been successfully
implemented for homogeneous electron systems, including hot plasmas and quantum Hall fluids.  
The method leads to positive $g(r)$ at all couplings and satisfies the known sum rules adequately.
We recall that Laughlin's plasma model for
the quantum Hall effect\cite{Laughlin83}, extended by Haldane\cite{Haldane1983}, Halperin\cite{Halperin1984},
MacDonald {\it et al.}\cite{MacDonald1985},  needs an ansatz wavefunction, and uses an effective quantum
 temperature for the classical fluid, even for quantum systems at zero temperature.  
The hyper-netted-chain (HNC) equation was used by Laughlin\cite{Laughlin83} 
to obtain the pair distribution functions of the quantum Hall fluid.  

The CHNC method exploits Density-Functional Theory (DFT) ideas based on a single determinantal wavefunction.
DFT uses a single-particle wavefunction with an exchange-correlation (XC) functional, even for many-body systems.  
In the CHNC method, the temperature of a classical Coulomb fluid is chosen to reproduce the XC-energy of the 
quantum fluid at zero temperature.   The pair distribution functions and local field factors of the 
electron fluid can then be calculated at arbitrary temperatures, densities, and spin polarizations
 using simple generalizations.
The resulting CHNC pair distribution functions and local field factors were shown to
 be in good  agreement, where comparable results are available, with results from 
QMC simulations for the 2D electron fluid\cite{prl2,LFC03}.
The method has been further successfully applied to multi-component  quantum electron layers and also 
to hydrogen plasmas, but no previous applications to double quantum wells or coupled layers have been presented. 

The linear density-density response function $\chi(q,\omega)$ for the 2D electron fluid depends on
many-body interactions, which in DFT are treated as exchange-correlation effects.   
As usual, we express the response function $\chi(q,\omega)$ 
in terms of a reference ``zeroth-order''  $\chi^0_R(q,\omega)$ and a local field factor,
 denoted by $G(q,\omega)$\cite{VigGiu05},
\begin{equation}
\label{lffdef}
\chi(q,\omega)=\chi^0_R(q,\omega)/[1-(2\pi/q)\{1-G(q,\omega)\}\chi^0_R(q,\omega)].
\end{equation}

In Eq.\ \ref{lffdef}, the usual 2D bare Coulomb potential $V(q)=2\pi/q$ is used.
The many-body effects are contained in the local field factor $G(q,\omega)$.
Note that in the random phase approximation XC-effects are neglected, so $G(q,\omega)=0$.
The local field factor is closely related to the vertex function $\Lambda(q,\omega)$ of the
 electron-hole propagator.
The static form of the local field factor, $G(q)$, is identical with $G(q,0)$.
Considerable efforts have been devoted to determining
 $G(q)$, using perturbation theory,  kinetic-equation methods\cite{rk,stls}, etc.
A partially analytic, semi-empirical approach invokes parametrized models constrained
to satisfy sum rules\cite{iwamoto} which are then fitted\cite{marinescu,teter} to limited results
obtained from QMC simulations\cite{moroni95, davoudi01}.
However such methods are not feasible at finite temperatures.

In the present study we determine temperature-dependent pair distribution functions
and the local field factors needed for understanding the properties of double quantum wells  at finite-$T$.  
We use the HNC equation rather than the more complicated Modified HNC equation (MHNC) for the
following reasons. The MHNC 
includes a ``bridge diagram contribution'', and
 improves the calculated pair distribution
 functions at strong coupling. However, as shown in Ref.~\onlinecite{LFC03}, the local field
 factors are already in very good  agreement with the QMC results when the HNC equation is
  used, while the available hard-disk ansatz for the bridge contributions\cite{prl2}
 provides no further improvement in the local field factors. This justifies our use
of the HNC instead of the MHNC equation.

\section{The Classical Map hyper-netted-chain technique}
\label{chnc-dw.sec}

We now outline the established CHNC method and our extension of the method to
 the double quantum well system.
The charge carriers are of two spin species, so in principle a double quantum well contains
$n_c=4$ (four) components,  requiring self-consistent evaluation of $n_c(n_c+1)/2=10$ (ten)
pair distribution functions.   However, for equal densities and spin-unpolarized carriers,
there are only two pair distribution functions which are different.
Thus an unpolarized two-component up- and down-spin electron
(or hole) layer can be reduced to an effective single-component paramagnetic
 fluid.  This transforms the problem into a two component problem
 with only three independent pair distribution functions.
 %, $g_{11}, g_{12}$ and $g_{22}$.  
% Since $g_{11}=g_{22}$ for paramagnetic systems, 
 % we effectively need to discuss just two  pair distribution functions.

\subsection{The Method}
\label{CHNC-sec}

The classical-map HNC approach for a single system of fermions (e.g., 3D fliud,
or a 2D layer)
was discussed in a number of 
papers\cite{prl1,prb00,LiuWuCHNC18, prl2,pd2d,BulTan02,Totsuji09}.
It was shown that the static properties of the 2D and 3D electron systems,
(as well as electron-proton systems\cite{hug02}), can be calculated
{\it via} an equivalent {\it classical} Coulomb fluid having an effective ``classical-fluid''
temperature  $T_{cf}$ such that the classical fluid has the same correlation energy as the
quantum system. The exchange energy is already exactly included
in the method, since the zeroth order pair distribution function is constructed from the
Slater determinant of the free-electron (or hole) fluid.  At  $T=0$, the corresponding $T_{cf}$ is called the
``quantum temperature'' $T_q$ and can be determined easily using the known
XC-energies of the uniform electron fluid. 

Once $T_q$ is set, the method can be used to determine
pair distribution functions, local field factors and XC-energies wherever QMC data are unavailable,
 as was the case for  finite-$T$ 3D systems. For instance, the finite-$T$ XC-energies
  for the 3D electron system
using the classical map HNC\cite{prb00}  given in the year 2000, agreed very well with
the QMC results which only became available more than a decade later\cite{Hungary16}.
Applications to many systems and to hot-dense plasmas are given in Refs.~\onlinecite{Bredow15,Hungary16}. 
It should also be noted that classical Molecular Dynamics (MD) simulations can be used to determine the
 pair distribution functions of the equivalent classical fluid. However, 
the HNC integral equation provides a computationally very efficient and adequately 
accurate method for uniform systems.

The mapping is based on a physically motivated extension of the classical
Kohn-Sham equation, i.e., a Boltzmann-like equation for the
density $n(r)=\exp\{-V_{KS}(r)/T_q\}$ at $T_q$ that mimics the quantum system.
The quantum temperature $T_q$ applies when the system is at
 the physical temperature $T=0$.  The 2D $T_q$ was fitted to the form\cite{prl2},
\begin{equation}
\label{2dmap}
t=T_q/E_F=2/[1+0.86413(r_s^{1/6}-1)^2]\ ,
\end{equation}
where $E_F=1/r_s^2$ is the electron Fermi energy in Hartrees, with $r_s$ the
average interparticle spacing within a layer.  $T_q$ is also in Hartrees.
(Effective atomic units which subsume the effective mass and the material
 dielectric constant are used throughout.)
Other possible improved forms for $T_q$ have been discussed by Totsuji {\it et al.}\cite{Totsuji09}, but they
lead to similar results as Eq.~\ref{2dmap} in the range of $r_s$ that is of interest to us in this study.  
At finite temperature $T$, the classical-fluid temperature $T_{cf}$ is taken to be
 $T_{cf}=(T_q^2+T^2)^{1/2}$, as discussed in Refs.~\onlinecite{prb00,SandipDufty13}.

In this section we discuss only a single layer or quantum well treated as an infinitely thin sheet. 
The extension to double quantum wells is given in Sec.~\ref{dw-sec}.  
The pair distribution functions are given by the HNC equation\cite{hncref} extended to include the  bridge
 terms (i.e., in effect, the MHNC equation).
The MHNC equations,  the Ornstein-Zernike relations for the pair distribution functions $g_{ij}(r)$,
and the ``direct correlation function'' $c_{ij}(r)$ are\cite{hncref}:
\begin{eqnarray}
\label{hnc1}
g_{ij}(r)&=&\exp[-\beta_{cf} \phi_{ij}(r)
+h_{ij}(r)-c_{ij}(r) + B_{ij}(r)]\nonumber\\
 h_{ij}(r) &=& c_{ij}(r)+
\Sigma_s \> n_s\! \int d{\bf r}'h_{i,s}
(|{\bf r}-{\bf r}'|)c_{s,j}({\bf r}') \ .
\label{hnc2}
\end{eqnarray}
The inverse temperature $\beta_{cf}=1/T_{cf}$.  The subscripts here denote the spin indices.
The total correlation function $h_{ij}(r)=g_{ij}(r)-1$ has been introduced.  
These relations involve: 
(i) the pair-potential $\phi_{ij}(r)$, and 
(ii) the bridge function $ B_{ij}(r)$\cite{rosen87}. 

When the bridge contribution (clusters beyond the hyper-netted-chain  diagrams) is set to zero
 we get the HNC equation.  If a classical  MD simulation is used to obtain the pair distribution
functions of the ``equivalent'' classical fluid, then the bridge term is automatically 
included without the need for hard-sphere models used in MHNC.  
The relevant pair-potentials $\phi_{ij}(r)$ for interacting particles are 
\begin{eqnarray}
\label{pair-pots.eqn}
\label{phi-ij.eqn}
\phi_{ij}(r)&=&{\cal{P}}(r)\delta_{ij}+V^c(r) \\
\label{Pauli.eqn}
{\cal{P}}(r)&=&h^0_{ii}(r)-c^0_{ii}(r)-\ln[g^0_{ii}(r)]\ .
\end{eqnarray}
${\cal{P}}(r)$  is the ``Pauli exclusion potential'' which brings in exchange effects contained
in the non-interacting pair-distribution function $g^0_{ii}(r)$. The Coulomb interaction between
a pair of particles is denoted by  $V^c(r)$. Since we are treating paramagnetic electrons $g^0_{11}=g^0_{22}$, so
we  have suppressed the spin indices on ${\cal{P}}$, except when needed for clarity.
In Sec.~\ref{dw-sec} we generalize these potentials $\phi_{ij}(r)$ for applications to double quantum wells.

The individual pair distribution functions $g_{ij}(r)$ depend on the pair-potentials $\phi_{ij}(r)$, 
as given in the HNC equations.
Eq.~\ref{pair-pots.eqn} treats the pair potentials as a sum of the Coulomb interaction $V^c(r)$ and
the  Pauli exclusion potential ${\cal{P}}(r)$.
The latter mimics the exchange hole arising from the antisymmetry of the underlying
Slater determinant, which is the only wavefunction used in DFT, even for 
many-particle systems.  Since the non-interacting $g^0_{ij}(r)$ do not contain
the Coulomb potential, the Pauli exclusion potential ${\cal{P}}(r)$ (which is in effect a
 kinematic interaction) can be obtained by an
inversion of $g^0_{ii}(r)$ via the HNC equation\cite{Lado67}, as summarized in Eq.~\ref{Pauli.eqn}.
Since $g^0_{12}(r)=1$, the Pauli potential  ${\cal{P}}_{12}(r)=0$  for
antiparallel spins.   The Pauli potential between two parallel-spin electrons is
obtained by HNC-inversion via Eq.~\ref{Pauli.eqn}.  This potential
is repulsive, long-ranged, and scale independent (i.e depends only on $r/r_s$).

\subsection{Reduction of the two-spin fluid  to a single effective fluid}
\label{reduce-sec}
In this study we consider only zero spin polarization, $\zeta=0$. Hence   
an averaged pair distribution function for the paramagnetic electron fluid in a
 single layer can be constructed, 
\begin{equation}
 g_{p}(r)=\{g_{11}(r)+g_{12}(r)\}/2 \ .
\end{equation}
Since the $g_p(r)$ is an average, the corresponding Pauli-exclusion potential
 ${\cal{P}}(r)$ is not the same as that used in $g_{11}(r)$,
 but needs to be determined anew, using $g^0_p(r)$ at the given density and   temperature as input.   
 The use of such an average potential  and an average $g_p(r)$ is justified as long as there are no
 magnetic or  spin-dependent interactions in the Hamiltonian.  
 The density of the carriers in the quantum well is the full carrier density $n$, 
 while for $\zeta=0$, the density of each  spin component  is $n/2$. 

The Coulomb potential used in the quantum problem is the {\it operator} $1/r$.
In the classical map, the potential is 
an effective  Coulomb potential $V^c(r)$ containing  a diffraction correction  associated
with the de Broglie wavelength  of the interacting electron pair at their
classical fluid  temperature $T_{cf}$.  It may be noted that this `regularization'
of the Coulomb potential for small $r$  is similar to the use of the Compton
cutoff momentum in high-energy collisions.  $T_{cf}$ defines 
the de Borglie thermal momentum  of the pair.
\begin{equation}
\label{kth.eqn}
k_{th}=\surd(2\pi m_r T_{cf}),\;\;\; m_r=m^\star/2.
\end{equation} 
For equal effective masses, the reduced mass is $m_r=m^\star/2$.
Improved forms of $k_{th}$ and $T_{cf}$ to be used in 2D  CHNC have been proposed by
Totsuji {\it et al.}\cite{Totsuji09}.  These are however not expected to
play a significant change for the range of $r_s$ and $T$ studied here, and hence
we use the original parametrizations given in Ref.~\onlinecite{prl2}.

For an interacting pair of carriers in a 2D layer we have,
\begin{eqnarray}
\label{Coul-eq}
V^c(r)&=& (1/r)\{1-e^{-k_{th}r}\} \nonumber \\
V^c(q)&=&2\pi \{1/q-1/(q^2+k^2_{th})^{1/2} \} \ .
\end{eqnarray}
The 2D-Fourier transform of $V^c(r)$ is denoted by $V^c(q)$.
As already noted, we use units $\hbar=e=m_e=1$, and effective atomic units
containing the effective mass $m^\star$ and the background dielectric
 constant $\kappa$ of the quantum well.
The classical Coulomb potential in Eq.~\ref{Coul-eq}, called a
 ``diffraction-corrected'' potential,
behaves as a Coulomb potential for length-scales larger
than a de Broglie wavelength $\sim 1/k_{th}$.  However, for close approach
 the potential is not singular and reduces to a finite value, viz., $k_{th}$.

We solve the HNC equations, Eqs.\ \ref{hnc2}, using an iterative  numerical approach similar
to that given by Ng\cite{Ng1974}. The essential point is to remove long-range
interactions coming from the Coulomb and Pauli potentials  and to
treat them analytically in doing the Fourier transforms, while the short-range parts
 have to be done numerically.
The Pauli potential and direct correlations functions
derived from the non-interacting $g^0(r)$ serve as the initial inputs to start off the
interations inclusive of the Coulomb interactions.  

For $0<t=T/E_F<1$ the electron system remains partially degenerate, while for $t>1$,
 the electrons will approach classical behavior. Classical correlations scale
according to the coupling parameter $\Gamma=1/(r_sT)$.  This contrasts with the quantum
 correlations at $T=0$ that scale with $r_s$.
When $T \gg T_q$, only classical correlations are important for $r>1/k_{th}$.
In the partially degenerate regime there is no simple coupling parameter, but in
 constructing our $\Gamma$, 
the classical fluid temperature $T_{cf}$ replaces $T$. 

\subsection{Calculation of the local field factors}
\label{LFC-sec}

The pair-distribution functions  $g_{ij}(r)$ can then be used
to extract the local field factors for the quantum fluid.  
The structure factor $S_{ij}(q)$ is related to the $g_{ij}(r)$
by the usual Fourier transform.  
In contrast to the quantum case, for a classical fluid the
density-density linear response function $\chi_{ij}(q)$ is
directly related to the structure factor,
\begin{equation}
S_{ij}(q)=-(1/\beta_{cf})\chi_{ij}(q)/n \ .
\label{StoChi}
\end{equation}
For the single well, the static local field $G(k)$ for the paramagnetic case is obtained from,
\begin{equation}
\label{lfcCS}
V^{c}(q)G(q)=V^{c}(q) -\frac{T_{cf}}{n} \Bigl\lbrack\frac{1}{S(q)}-\frac{1}{S^0(q)}\Bigr\rbrack .
\end{equation}

In CHNC, the structure factor for the non-interacting system, $S^0(q)$, is based on a
Slater determinant and not on the non-interacting structure factor
corresponding to the Lindhard function $\chi_L^0$.
QMC results use a reference $\chi_L^0$ such that the local field factor contains a
 kinetic-energy tail, as discussed in Ref.~\onlinecite{LFC03}.
 The $S^0(q)$ for the non-interacting 2D electron fluid
 is numerically known at any $T$, and hence the calculation
of the temperature-dependent local field factor is simple, once the interacting $S(q)$ and the classical
temperature $T_{cf}$ are obtained from CHNC. 

For numerical work it is convenient to re-express the equation for the local field factor
in terms of the direct correlation functions $c_{ij}(k)$ using the following standard relations
among structure factors and direct correlation functions,
\begin{eqnarray}
\label{skhk=eq}
S_{ij}(q)&=& \delta_{ij}+ n h_{ij}(q)\nonumber \\
h_{ij}(q)&=& c_{ij}+\Sigma_{s}c_{is}n h_{sj}(q).
\end{eqnarray}
Then  it can be shown that, 
\begin{equation}
\label{LFF-eq}
G_{ij}(q)=\{\tilde{c}_{ij}(q)-c^0_{ij}(q)\}/ \{\beta_{cf} V^c(q)\} \ ,
\end{equation}
where $\tilde{c}_{ij}(q)=c_{ij}(q)+\beta_{cf} V^c(q)$ is the short-ranged direct correlation function.
The local field factor of the averaged paramagnetic fluid is given by:
\begin{equation}
\label{LFFpara-eq}
G_p(q)=\{G_{11}(q)+G_{12}(q)\}/2 \ ,
\end{equation}
where the  contributions from the two spin species in the single layer are explicitly displayed. 
The good agreement of local field factors for single layers  at $T=0$ obtained by these methods and
from QMC was presented in Ref.~\onlinecite{LFC03}.  Finite-$T$ local field factors
 are as yet not available from QMC or path-integral simulations of 2D layers.

\section{The CHNC method for double quantum wells}
\label{dw-sec}

We now generalize the discussion to two coupled layers (e.g., as in graphene)
or two coupled quantum wells. 
Our system consists of left and right wells 
separated by a barrier of width $b$.   The wells  are assumed to be infinitely
thin, so the barrier width $b$ should include the actual width of the barrier $b^0$,
 plus one-half of the widths $W$ of each well. 
Since $W$ is the same for symmetric wells, $b=b^0+W$. 
The barrier material, with only a few percent of Al in the GaAlAs alloy, is usually 
not too different from the well material (GaAs), so we take the static dielectric constant
of the barrier to be the same as that of the well. 

As we are working with paramagnetic fluids and their pair distribution functions $g_p(r)$ 
with appropriate exchange interaction, there are no longer any spin indices.
Therefore from now on, we use  indices to refer to the left (1) and right (2) wells. 
The  intralayer Coulomb interaction $V^c(r)$ (Eq.~\ref{Coul-eq}), is now written
 $V^c_{11}(r)=V^c_{22}(r)$, while the interlayer interaction between carriers
 across the barrier  is  $V_{12}(r)$.   

It is conventional to approximate the interlayer interaction  by: 
  \begin{eqnarray}
V_{12}(r)&=&z_1z_2/\rho;\;\;\rho=\surd(r^2+b^2) \label{Coul-opp(r)}\\
V_{12}(q)&=&2\pi z_1z_2 e^{-bq}/q \ .  \label{Coul-opp(q)}
\end{eqnarray}
$z_i=\mp 1$ is the charge of the carriers in layer $i$, and $r$ is the in-plane distance.  
Since carriers in opposite layers are distinct fermions, in the classical map there is
no Pauli-exclusion potential acting between left and right layer carriers.  
The interlayer  interaction acts on the electron wavefunctions to produce
 a modified Coulomb potential.  Equation \ref{Coul-opp(r)} is thus only approximately
true for close approach. The form we adopt for the diffraction-corrected
classical Coulomb potential across the barrier is:
\begin{equation}
\label{Coul-eq12}
V^c_{12}(r)=z_1z_2\{1-e^{-k^b_{th}\rho}\}/\rho \ .
\end{equation} 

There is some ambiguity here in the choice of the thermal cutoff wave vector 
$k^b_{th}$ when the interacting pair is also separated by the barrier thickness $b$,
even for symmetric double wells at  the same temperature and density. In each layer
$k_{th}$ corresponds to a de Broglie length $\lambda=2\pi/k_{th}$. When acting
across the barrier, we include the effect of the barrier width as well
 in limiting close encounters, and use $\lambda^b=b+\lambda$. Then $k^b_{th}=2\pi/\lambda^b$.
 This  correction is of any importance only for {\it e-h} pairs where the Coulomb potential is
 attractive and the treatment of short-ranged interactions is of importance. In
effect, infintely thin layers are not acceptable in a consistent classical
map since the classical potential is meaningful only if the layers have a
minimal thickness that can support at least half a de Broglie wavelength. 
However, in order to maintain the transparency of the computation, in this
study we have retained the approximation of using a common $k_{th}$ everywhere.
 Instead, if $k^b_{th}=2\pi/\lambda^b$ were used, the value of
$g_{eh}(r)$ as $r\to 0$ is reduced somewhat, especially for larger $r_s$.
QMC benchmarks and alternative calculations would be very useful in
clarifying the accuracy of such approximations.
 
Furthermore, if the classical temperatures $T_{cf}$ of the layers were different,
as with asymmetric systems, then further considerations are needed. 
Then it can be shown that a good approximation is  to use the geometric mean of the
thermal $k_{th}$ of the two components in the above approach.  This has been
tested for  3D CHNC calculations for the two components having
different temperatures\cite{Bredow15}.

%\subsection{Calculation of the local field factors}
%\label{lff-sec}

%
\begin{figure}[t]
\includegraphics[width=1\columnwidth]{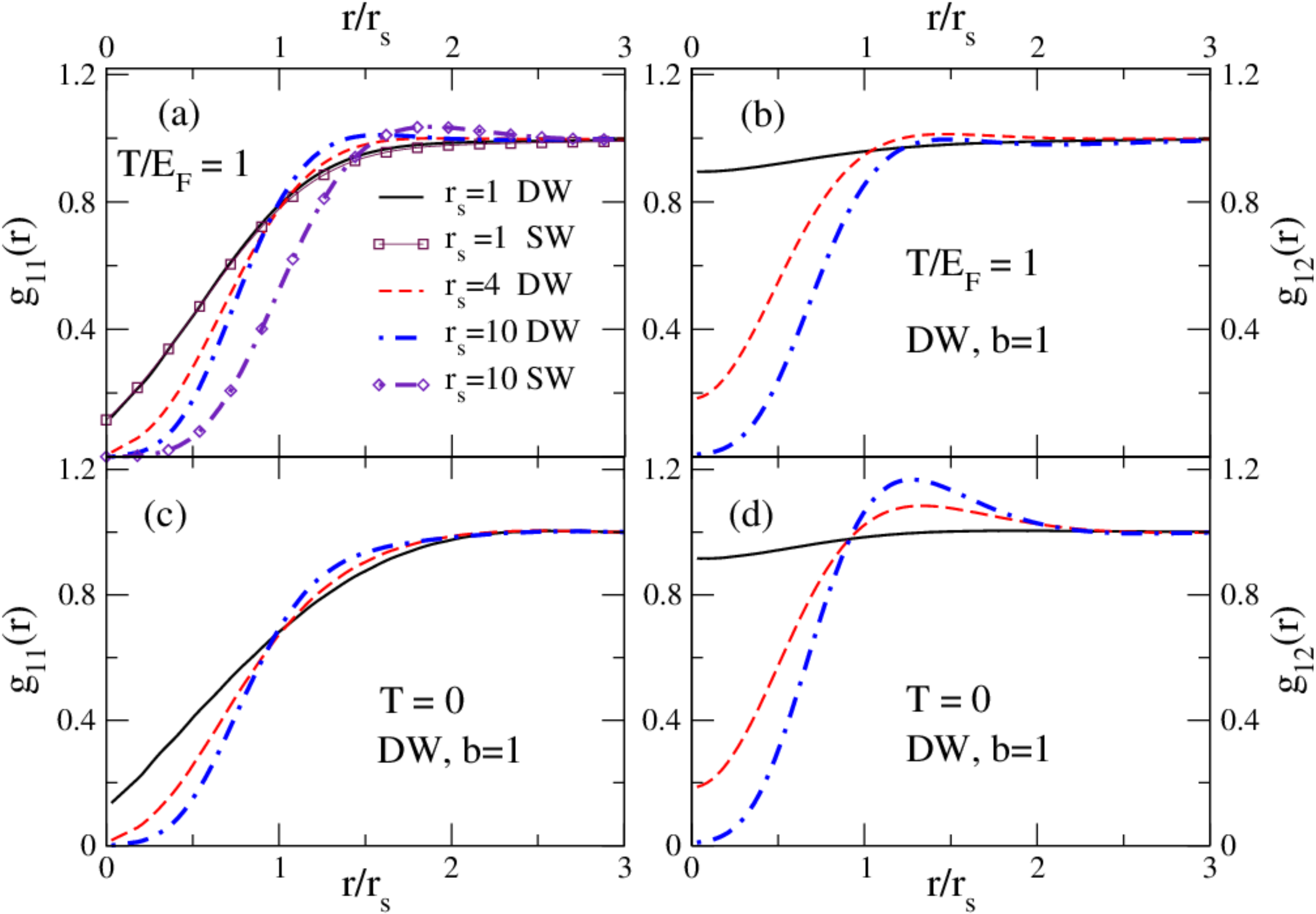}         % Fig_1
\caption{(Color online) Panels (a),(b) show the pair distribution functions $g_{11}(r),g_{12}(r)$ for
 paramagnetic electrons in a double well (DW) of separation $b=1$, at fixed finite temperature $T/E_F=1$, for  
densities $r_s=1,4,10$. The paramagnetic $g_p(r)$ for a single well (SW) at $T/E_F=1$ is
also shown for $r_s=1$ and $10$. Panels (c),(d) display the corresponding pair distribution functions at $T=0$  
for the double well only.
}
\label{pdfrs.fig}
\end{figure}
\begin{figure}[t]
\includegraphics[width=1\columnwidth]{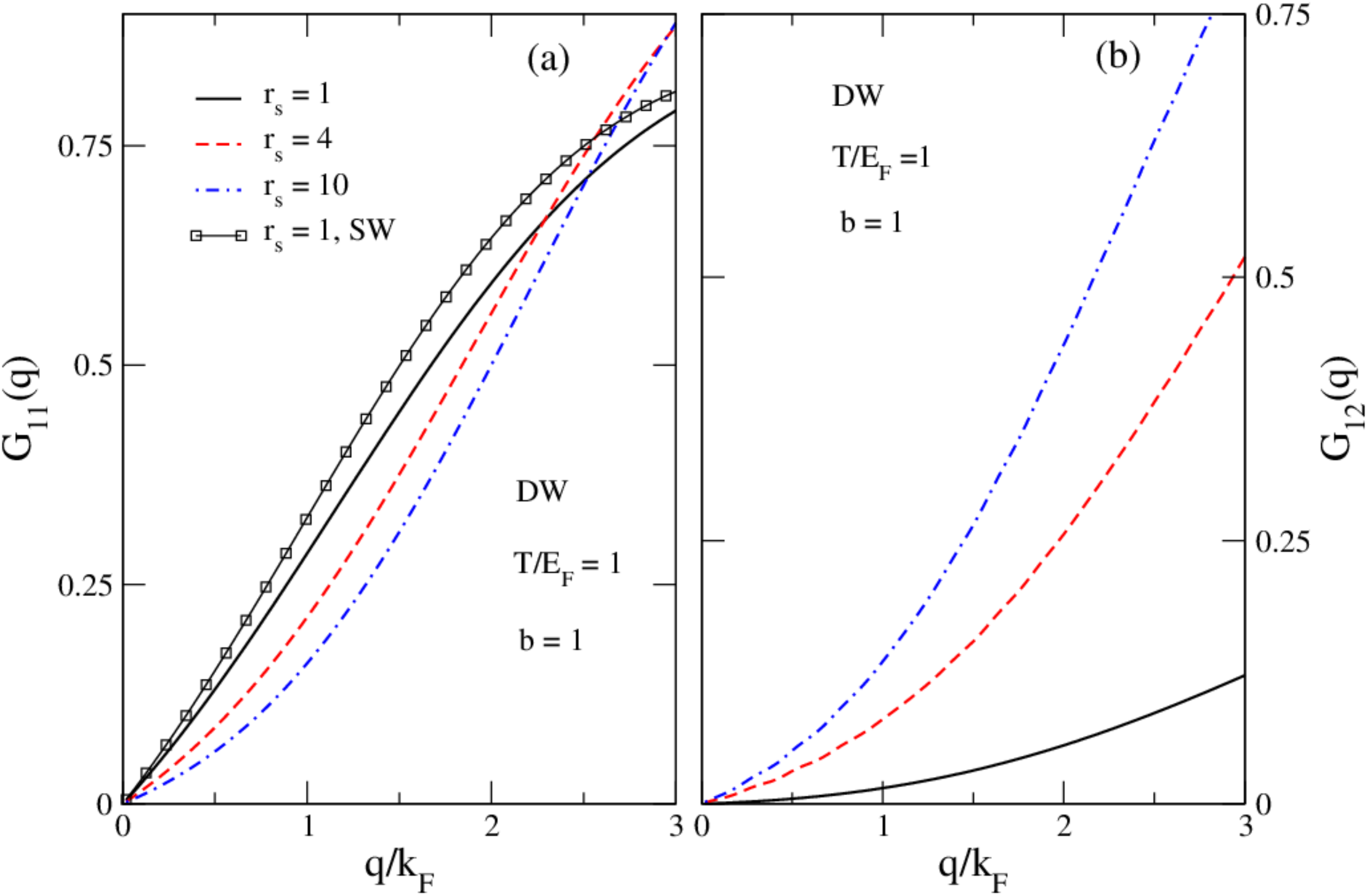}          %Fig_2
\caption{(Color online) (a) The intralayer static local field factors $G_{11}(q)$ for paramagnetic electrons 
in a double quantum well (DW) of separation $b=1$, at temperature $T/E_F=1$ 
for different densities  $r_s=1,4,10$. 
The  $G_{11}(q)$ for a single well (SW) at $T/E_F=1$ is also shown for $r_s=1$. 
(b) The corresponding interlayer static local field factors $G_{12}(q)$ for the double well only. 
}
\label{LFFrs.fig}
\end{figure}
The intralayer local field factors $G_{11}(q)= G_{22}(q)$
and the interlayer local field factor $G_{12}(q)$ are determined for the double quantum wells in analogy to 
Eq.~\ref{LFF-eq}, but with the indices now referring to the layers, and using  the appropriate diffraction-corrected 
Coulomb potentials $V^c_{ij}(q)$ (see Eqs.\ \ref{Coul-eq} and \ref{Coul-eq12}), 
\begin{equation}
\label{LFF2-eq}
G_{ij}(q)=\{\tilde{c}_{ij}(q)-c^0_{ij}(q)\}/ \{\beta_{cf} V^c_{ij}(q)\} \ ,
\end{equation}
with $\tilde{c}_{ij}(q)=c_{ij}(q)+\beta_{cf} V^c_{ij}(q)$. 
\begin{figure}[t]
\includegraphics[width=1\columnwidth]{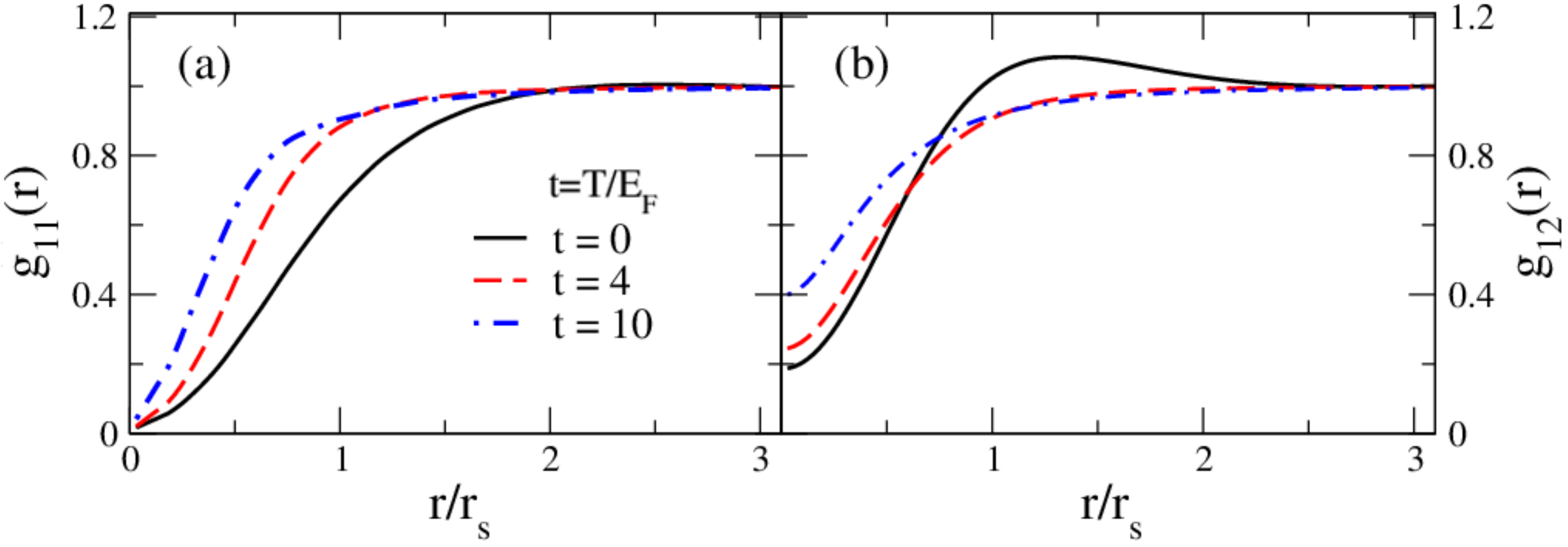}      %Fig_3
\caption{(Color online) The interlayer and intralayer pair distribution functions
 for  paramagnetic electrons in a double quantum well of separation $b=1$, for
 temperatures $t=T/E_F$=0,4,10 at a fixed density $r_s=4$. 
 }
\label{gr-T.fig}
%Figure 3
\end{figure}

\section{Double quantum wells with carriers of identical charge and mass.}
\label{eepair distribution functions-sec}

We present results for symmetric double quantum wells containing the same
 unpolarized carriers at finite temperature $t=T/E_F$ and (equal)
 average interparticle spacings,
 $r_s$, within the  layers.  Here we take two wells separated by a barrier of width $b=1$ 
 (corresponding  to $\sim 5$ nm in graphene and $\sim 10$ nm in  GaAs).
Finite temperatures can be accessed using Feynman-Path
 integral methods, and such results would be valuable for bench-marking the
CHNC results.  However, no calculations are so far available for this system.

\begin{figure}[t]
\includegraphics[width=1\columnwidth]{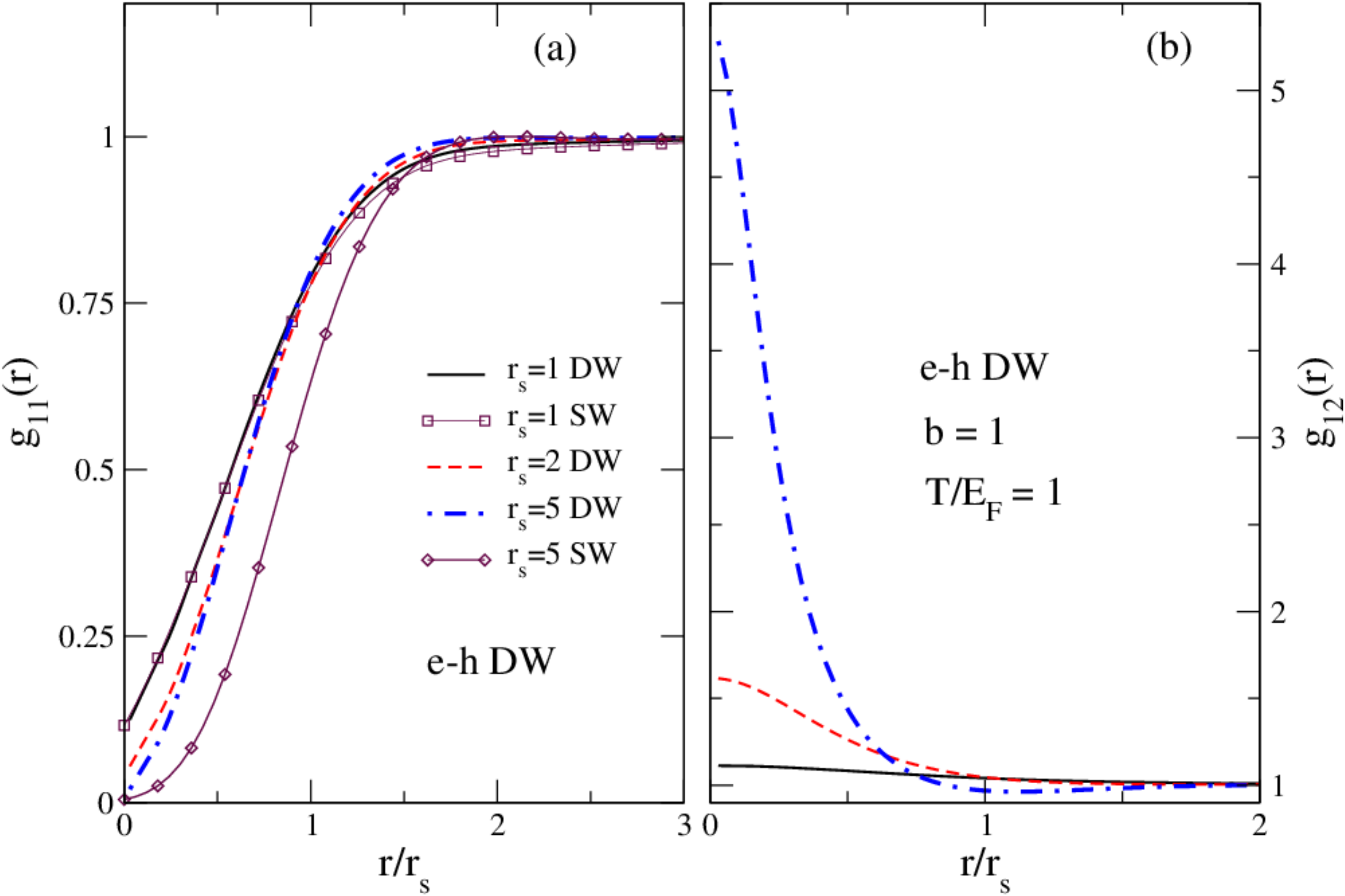}         % Fig_4
\caption{(Color online)  Variation of the intralayer and  interlayer pair-distribution functions $g_{11}(r)$
 and $g_{12}(r)$ at fixed finite temperature $T/E_F=1$, for  
electron-hole quantum double wells (DW) separated by a barrier of thickness $b=1$, for different densities $r_s = 1,4,5$. 
Comparison of $g_{11}(r)$ with the $g_p(r)$ of a single well (SW) is also given in (a) for $r_s=1$ and 5. 
The value of $\lim_{r\to 0}g_{12}(r)$ increases dramatically with increased coupling (larger $r_s$).
}
\label{pdfrsEH.fig}
\end{figure}
\begin{figure}[t]
\includegraphics[width=1\columnwidth]{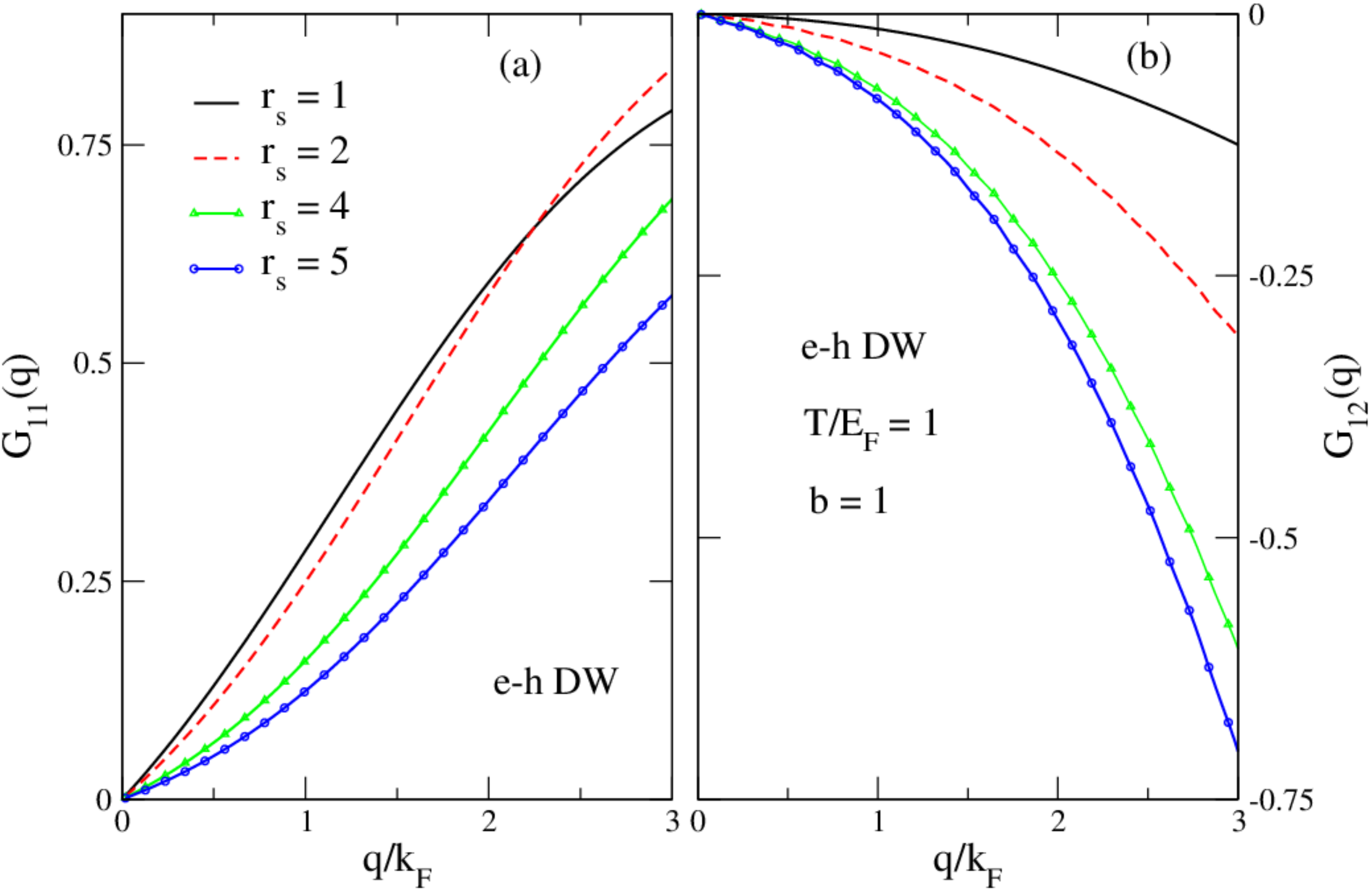}      % Fig_5
\caption{(Color online) The intralayer and interlayer local field factors $G_{11}(q)$ and $G_{12}(q)$ 
 for paramagnetic fluids in electron-hole double quantum wells of separation $b=1$, at fixed
 finite temperature $T/E_F=1$. 
Results for the densities corresponding to $r_s = 1, 2, 4, 5$, are displayed.   
The interlayer local field factor $G_{12}(q)$ becomes negative in the electron-hole system.
}
\label{LFFrsEH.fig}
\end{figure}
In Figs.~\ref{pdfrs.fig}(a) and (c) we display the intralayer pair distribution functions $g_{11}(r)=g_{22}(r)$
for two layers at fixed temperatures $T/E_F=1$ and $T=0$, for carrier densities with $r_s=1$ to 10.
In electrons in GaAs wells, this range corresponds to densities of $n \simeq 3\times 10^{11}$ to
 $3\times 10^{9}$ cm$^{-2}$.  
For this density range we see  that the in-layer pair distribution functions $g_{11}(r)$ are not
 very sensitive to changes in temperature, 
at least up to $T/E_F=1$.

At the high density $r_s=1$,  $g_{11}(r)$ for the  double quantum well  (black line) is almost
identical to the paramagnetic $g_p(r)$ of a single quantum well (curve marked with boxes).   
 However with decreasing density, as the Coulomb-interaction energy becomes relatively
 stronger compared to the Fermi energy
 (e.g., for $r_s=10$), we see that the pair distribution functions
 for the double and single wells are substantially different.
The double quantum well pair distribution function $g_{11}(r)$ is less strongly coupled than in a single well, 
with its maximum at a smaller $r/r_s$. 
For lower densities, the double quantum well $g_{11}(r)$ behaves in a manner similar to $g_p(r)$ 
 of a single well at nearly twice the density.  
This is to be expected at densities for which the average interparticle spacing in a well 
is much larger than the barrier separation, $r_s\gg b$. However, this implies that using local field
factors calculated for single wells for use in double well studies,
can become  a significant source of error for larger  $r_s$ values. 

In Figs.~\ref{pdfrs.fig}(b) and (d) we display the corresponding  interlayer pair distribution
 functions $g_{12}(r)$, which are a measure of the Coulomb correlations between the layers.  
While the very short-range interlayer correlations are only weakly affected by temperature, at least
 up to $T/E_F=1$ for the density range considered,
at larger $r/r_s$ the peak found at lower densities in the zero temperature $g_{12}(r)$ that
is centered near $r/r_s=1.25$,  
is already completely suppressed by $T/E_F=1$.     
Interestingly, the peak height in $g_{12}(r)$ grows until about $r_s\sim 6$, after which it 
decreases slightly for higher $r_s$ values.   
This is further evidence that, as the Coulomb coupling becomes more important relative to the kinetic energy, 
the quantum double well behaves increasingly like a single, wider well with larger effective density.

Local field factors at finite $T/E_F=1$ are displayed in Figs.~\ref{LFFrs.fig}(a) and (b) for the
 density range corresponding to $r_s=1$ to $10$.   
The intralayer local field factor $G_{11}(q)$ is only weakly dependent on density, but the interlayer
local field factor $G_{12}(q)$,
which is small for $r_s=1$, grows with decreasing density, and by $r_s=10$ it has approached the form of $G_{11}(q)$.
This is another indication that the barrier separation, fixed here at $b=1$, has become so small compared with 
the average interparticle spacing that the separation of the layers no longer affects the correlations. 
The changes in $G_{12}(q)$  with $r_s$ are large by $q=2k_F$, which is the important $q$-vector range for
the interactions. 

Figure \ref{gr-T.fig} shows the pair distribution functions over a wider range of temperatures $t$.  
The barrier width is again $b=1$.  The density is fixed at $r_s=4$, 
corresponding to $n\simeq 2\times 10^{10}$  cm$^{-2}$ for electrons in a GaAs well.  
Both the intralayer and interlayer correlations become weaker with increasing $t$.   
We saw in Fig.\ \ref{pdfrs.fig} that the zero-temperature peak in $g_{12}(r)$ had already completely 
disappeared by $t=1$. 

\section{Electron-hole double quantum wells}   %******************
\label{eh.sec}

\begin{figure}[t]   % fig. 6
\begin{center}
\subfigure{\label{fig:edge-a}\includegraphics[width=0.49\columnwidth,height=5.1cm]{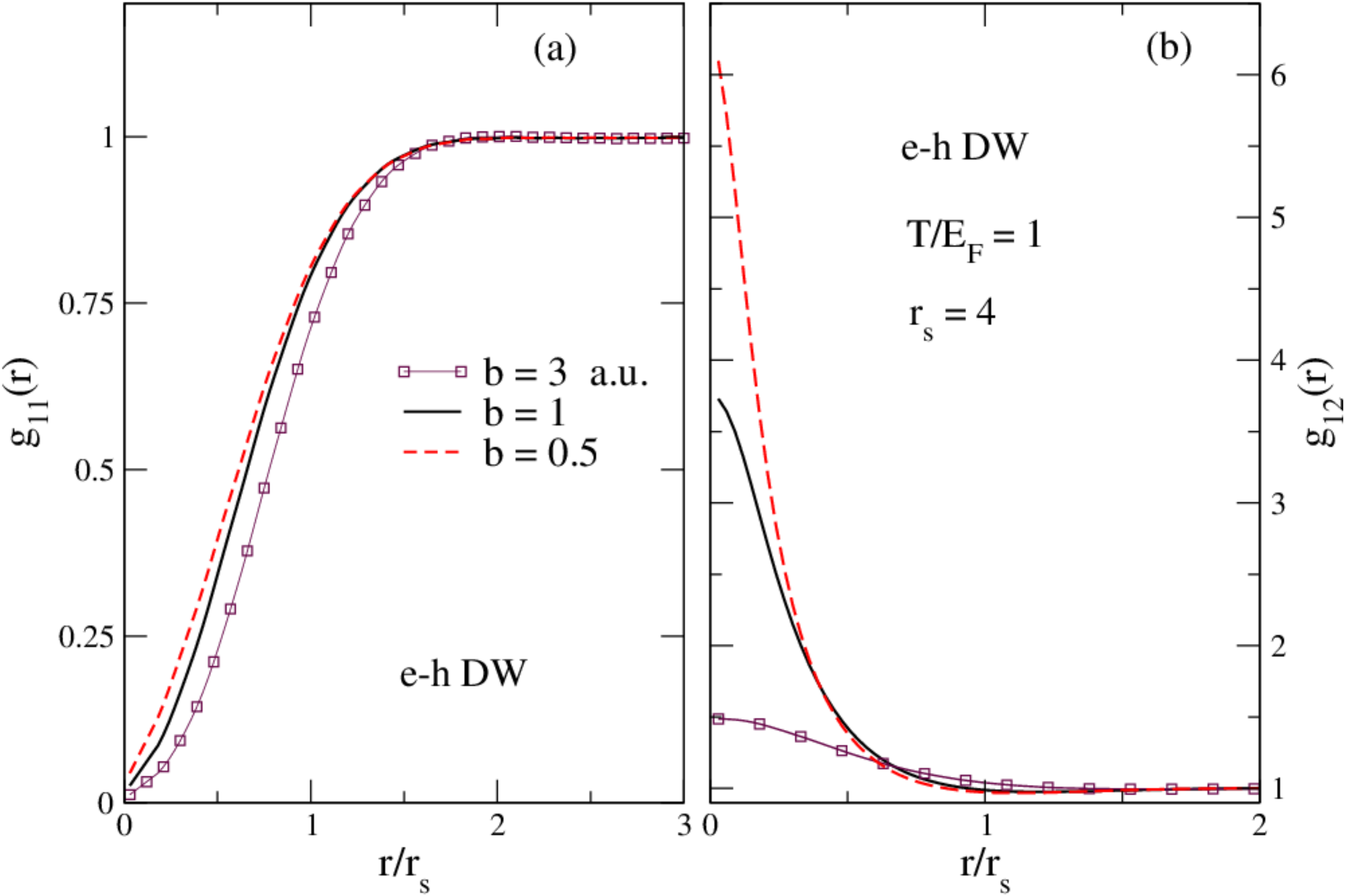}}
\subfigure{\label{fig:edge-b}\includegraphics[width=0.49\columnwidth,height=5.15cm]{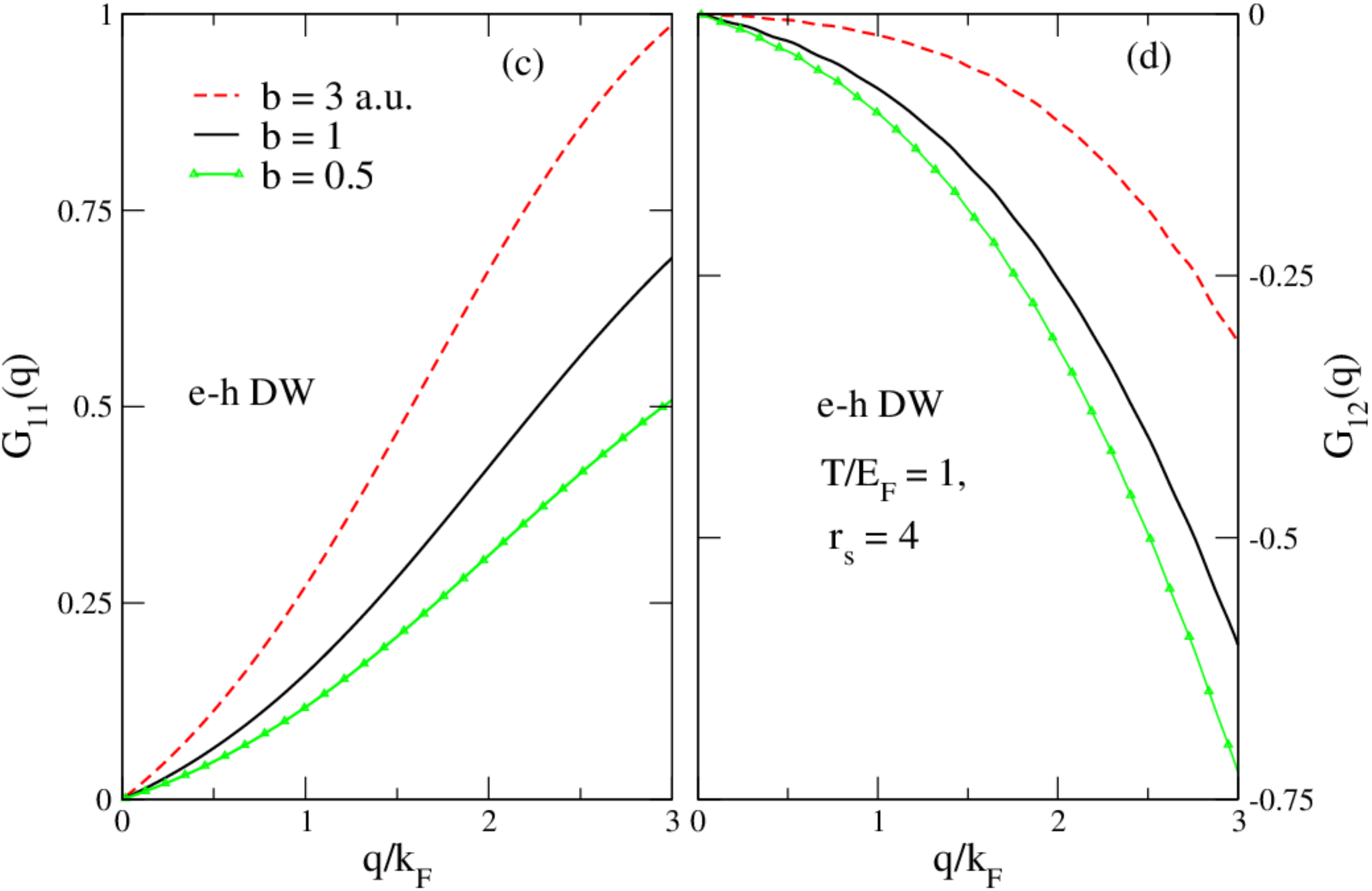}}
\end{center}
\caption{(Color online) (a) The intralayer pair distribution functions for  
 paramagnetic fluids of fixed equal density $r_s=4$, in electron-hole double
 quantum wells at finite temperature $T/E_F=1$, 
for different barrier widths $b$, as labeled.    
(b) The corresponding interlayer pair distribution functions $g_{12}(r)$.  
As expected, the value of $g_{12}(r)$ decreases with increasing $b$. 
(c) The corresponding intralayer local field factors  $G_{11}(q)$.   
(d) The interlayer local field factors $G_{12}(q)$.
}
\label{gr-bar.fig}
\end{figure}

Figure \ref{pdfrsEH.fig}(a) shows the intralayer pair distribution functions for electron-hole layers 
at equal densities for fixed finite temperature $T/E_F=1$. As already noted,
the properties of symmetic double quantum wells  are of importance in applications to
 graphene-like systems where atomically
thin  (and essentailly equivalent) carrier layers are used. Furthermore, they remain the least
 challanging
candidate for future work using Quantum Monte Carlo and related first principles methods.

In Fig.~\ref{pdfrsEH.fig}  the barrier thickness is $b=1$.
For symmetric wells, $g_{11}(r)=g_{22}(r)$.  
Figure \ref{pdfrsEH.fig}(b) shows the corresponding  interlayer pair distribution  function $g_{12}(r)$.
The value of $\lim_{r\to 0} g_{12}(r)$ increases
with increased coupling (increasing $r_s$).  It should be appreciated that the
present theory goes completely beyond the usual mean-field theories that originated with
 Keldysh \cite{Keldysh65} and other early workers
 (as reviewed in, e.g., Ref. \onlinecite{Littlewood1996}). The CHNC is
 designed to include
XC-effects  arising from the interactions beyond mean-field effects.
%However, there is no convergence of the CHNC equations for $r_s\gtrsim5$ at lower
% densities where  electron-hole bound state formation occurrs.  
There is of course no
provision for excitonic states in the existing CHNC theory although the
calculation may remain  robust into the weakly bound excitonic regime before
it fails. 
However, the pairing of oppositely charged particles leads
to a rearrangement of the ground state of the system. This
is accompanied by the appearance of an order parameter
 proportional to the magnitude of the gap in the single- 
particle excitation spectrum of the system \cite{Lozovik1976a}.
Since the classical-map technique uses a $T_q$ (Eqn.~\ref{2dmap}) fitted to reproduce
the XC-energy of a simple Fermi liquid, the added correlations due to
pairing are not included in the present formulation. Since the method 
is motivated by density-functional ideas (e.g.., uses pair-densities instead of
 wavefunctions), the possibility of extending the method to regimes of exciton
 formation,  superfluidity etc., may perhaps  be envisaged, borrowing ideas
 from the   density-functional approach to
phenomena like  superconductivity~\cite{OliveiraGrossKohn88}

In Fig.~\ref{LFFrsEH.fig} we display the density dependence of the corresponding
 local field factors for  temperature $T/E_F=1$.   
 While the intralayer local field factors $G_{11}(q)$ are similar to those
 of electron-electron double quantum wells, a notable feature here is the
 negativity of the electron-hole interlayer local field factor $G_{12}(q)$.
   It is this feature that leads 
to zeroes in the denominators of response functions, signaling the formation of new 
elementary excitations, i.e., excitons in this case.

\subsection{The effect of the barrier width}
\label{barrier-sec}

The effect of increasing barrier thickness $b$ on the pair distribution functions
 and the local field factors is presented  in Figs.~\ref{gr-bar.fig} (a) to (d) for an
electron-hole double quantum well of fixed equal densities $r_s=4$.
The barrier width  $b$ is varied from $0.5$ to $3$,  corresponding in n-GaAs to a
 range
 from $5$ nm to $33$ nm. 
As expected, a thicker barrier weakens the coupling between
the layers, so $g_{12}(r)$ and $G_{12}(q)$ are proportionately weakened. 

We note the rapid rise of $\lim_{r\to 0} g_{12}(r)$ in Fig.~\ref{gr-bar.fig}(b) as
 the barrier width is diminished.
For density $r_s=4$, no convergence was obtained for barrier thickness less 
than $b\sim 0.267$, at which point $g_{12}(r=0)$ has reached $8.2$.  
In n-GaAs, $b\sim 0.267$ corresponds to a barrier thickness of $\sim 2.9$ nm.  
This lack of convergence is a consequence of very strong interactions that cause
excitonic bound-states to emerge in the physical system. On general grounds one may
expect that if the barrier thickness $b$ were greater than the mean exciton radius,
 then
the system may be relaibaly studied by the present formulation.

\section{Conclusions}

We have presented  results for the pair distribution functions and local field factors as a function
of temperature, density and barrier width for electron-electron and electron-hole double quantum wells. 
While the single-layer CHNC results have been checked against corresponding
 Quantum Monte Carlo results at $T=0$ to establish its accuracy, comparable
 QMC results are not yet available at finite-$T$. 
 
Our results confirm that there are  significant modifications  of
 the distribution functions and local field 
 factors due to finite-temperature effects, in particular when $T$ exceeds the Fermi temperature.
As already noted, in GaAs the Fermi temperature is only $4$ K at a hole layer
 density $n=4\times 10^{10}$ cm$^{-2}$. 
Our results also reveal that the local field factors calculated for single wells
cannot be used for the accurate calculations of properties of double quantum wells, 
unless the densities are high ($r_s\sim 1$).

The  local field factors with their density and temperature variation,  
need to be included in the linear
response functions that enter into many measurable properties of double
quantum wells. Such properties include (i) thermodynamic functions, (ii)
the drag resistivity of interacting double
 layers as a function of density, temperature and carrier type, (iii) plasmon dispersion
 in such layers  as a function of the density and temperature of the layers, and  
 (iv) energy relaxation of hot  electrons injected into one of the layers.  
 The CHNC formalism presented here can be readily generalized to spin polarized layers 
 and to layers with carriers of different effective masses.  Our formalism provides
 high computational efficiency, while providing  good to at least modest accuracy
 in regimes of strong correlations and finite temperatures where
 other methods become prohibitive.  

\begin{acknowledgments}
This work was partially supported by the Flemish Science Foundation (FWO-Vl).
C.D-W acknowledges with thanks the hospitality and stimulating atmosphere of the
 Condensed Matter Theory group at the University of Antwerp.
\end{acknowledgments}

\end{document}